\documentclass[11pt,a4paper]{article}
\usepackage[T1]{fontenc}
\usepackage{bm}  
\usepackage{amssymb}

\def\onehalf{\textstyle{\frac{1}{2}}}

\def\onefourth{\textstyle{\frac{1}{4}}}

\def\be{\begin{equation}}
\def\ee{\end{equation}}
\def\ba{\begin{eqnarray}}
\def\ea{\end{eqnarray}}

\begin{document}

\renewcommand{\thefootnote}{\fnsymbol{footnote}}
\begin{center}
{\Large \bf Spin-2 fields and helicity}
\vskip 0.5cm
{\bf H. I. Arcos$^{a,b}$, C. S. O. Mayor$^b$, G. Ot\'alora$^b$ and J. G. Pereira$^b$}
\vskip 0.3cm
$^a${\it Universidad Tecnol\'ogica de Pereira\\
A.A. 97, La Julita, Pereira, Colombia}
\vskip 0.2cm
$^b${\it Instituto de F\'{\i}sica Te\'orica, UNESP-Univ Estadual Paulista \\
Caixa Postal 70532-2, 01156-970 S\~ao Paulo, Brazil}

\vskip 0.8cm
\begin{quote}
{\bf Abstract.~}{\footnotesize By considering the irreducible representations of the Lorentz group, an analysis of the different spin-2 waves is presented. In particular, the question of the helicity is discussed. It is concluded that, although from the point of view of representation theory there are no compelling reasons to choose between spin-2 waves with helicity $\sigma = \pm 1$ or $\sigma = \pm 2$, consistency arguments of the ensuing field theories favor waves with helicity $\sigma = \pm 1$.

}
\end{quote}
\end{center}
\vskip 0.8cm 

%
%
%
%
%
%

\section{Introduction}
\label{intro}

For many years in the past, the existence (or not) of gravitational waves was controversial.  Einstein himself was not sure about their existence. Another interesting example is N. Rosen who published a paper in 1979 entitled {\it Does Gravitational Radiation Exist?}~\cite{rosen}. The discovery of a binary pulsar whose orbital period changes in accordance with the predicted gravitational wave emission \cite{HT} put an end to that controversy. In fact, that discovery provided compelling evidence for the existence of gravitational waves \cite{maggiore}. That evidence, however, did not provide any clue about their form and effects. It only confirmed the quadrupole radiation formula. Yet the traditional linear approach to gravitational wave theory became considered a finished topic, not to be questioned any more~\cite{giants}.

According to this approach, the dynamics of linear gravity is believed to coincide with the dynamics of a fundamental spin-2 field.  For this reason, gravitational waves became synonymous with spin-2 waves. Furthermore, since the spin of massless fields is usually said to be just the helicity of the field, gravitational waves are also believed to be waves with helicity $\sigma = \pm 2$. However, this analysis presents several obscure points \cite{cs1}. For example, though the Lorentz group has well--defined spin-2 representations with helicity $\sigma = \pm 1$, these waves are usually neglected in favor of waves with helicity $\sigma = \pm 2$. Is this assumption justified?

Through an analysis of all spin-2 representations of the Lorentz group, as well as the corresponding field equations, the purpose of this paper is to discuss this question, and possibly emerge with a bias towards  $\sigma = \pm 1$.

\section{Representations of the Lorentz group}
\label{Lorentz}

For every field (or particle) of nature there exists a representation of the Poincar\'e group \cite{repre}, the semi-direct product between Lorentz and the translation groups. The eigenvalues of the translational generators define the mass of the field, whereas the eigenvalues of the Lorentz generators define the spin of the field. The commutation relations of the six Lorentz generators, which we denote here by ${\mathbf a}$ and ${\mathbf b}$, can be written in the form \cite{weinberg}
\ba
{\bm a} \times {\bm a} = i {\bm a} \\
{\bm b} \times {\bm b} = i {\bm b} \, \\
\left[ a_i, b_j \right] = 0,
\ea
with $i, j, k, \dots = 1, 2, 3$. A general spin ${\sf s}$ representation of the Lorentz group can be constructed as either a field transforming under an irreducible representation, or as a direct sum of irreducible representations, each characterized by an integer or half-integer $A$ and $B$, with
\be
{\bm a}^2 = A(A +1) \quad \mbox{and} \quad {\bm b}^2 = B(B + 1).
\ee
These representations are labeled by the numbers $(A,B)$, where ${\sf s} = A+B$. The number of components $n$ of the representation $(A,B)$ is
\be
n = (2A+1)(2B+1).
\ee
For massless particles, which is the case we will be interested here,
\be
\sigma = B - A
\label{spinres}
\ee
represents the helicity of the corresponding wave \cite{weinbergf}.

When $A \neq B$, the irreducible representation can be written as a direct sum of the form $(A,B)\oplus(B,A)$, with the number of independent components given by $2 n$. The simplest example is the spin-0 field $\phi$, which is associated with the one-component (singlet) representation $(0,0)$. It satisfies the field equation\footnote{We use the Greek alphabet $\mu, \nu, \rho, \dots = 0,1,2,3$ to denote indices related to spacetime, also known as world indices. The first half of the Latin alphabet $a,b,c, \dots = 0,1,2,3$ will be used to denote algebraic indices related to the tangent spaces, each one a Minkowski spacetime with metric $\eta_{ab} = \mathrm{diag}(+1,-1,-1,-1)$.}
\be
\Box \, \phi = 0,
\ee
with $\Box = \eta^{\mu \nu} \, \partial_\mu \partial_\nu$ the d'Alembertian operator. The most fundamental representations, however, are the spinor representations. The $(1/2,0)$ representation has spin 1/2 and describes a (let us say) left--handed Weyl spinor.  The $(0,1/2)$ representation describes a right--handed Weyl spinor. The linear combination
\be
(1/2,0) \oplus (0,1/2)
\ee
describes a Dirac spinor. The reason why the spinor representations are the most fundamental is that they can be used to construct, by multiplying them together, any other representation of the Lorentz group.

\section{A warming up example: the massless spin-1 field}

In order to get some insight on the theory of representations, as well as on the corresponding wave equations, let us review the well--known case of a massless vector field.

\subsection{Potential waves}

The first way to construct a spin-1 representation is to consider the direct product
\be
(1/2,0) \otimes (0,1/2) = (1/2,1/2),
\ee
where $(1/2,1/2)$ describes a spin-1 field with $n=4$ components. The electromagnetic vector potential $A_\mu$ transforms according to this representation. In the Lorenz gauge
\be
\partial^\mu A_\mu = 0,
\label{EleLoreGau}
\ee
it satisfies the wave equation
\be
\Box A_{\mu} = 0.
\label{*}
\ee
The solutions to this equation represent, according to the definition~(\ref{spinres}), waves with helicity $\sigma = 0$.

\subsection{Field strength waves}

A second way to construct a spin-1 representation follows from the direct product
\be
(1/2,0) \otimes (1/2,0) = (1,0) \oplus (0,0).
\ee
The representation $(1,0)$ can be identified with an antisymmetric, self--dual second--rank tensor. Analogously, the representation $(0,1)$ can be identified with an antisymmetric, anti self--dual second--rank tensor. The representation that describes a parity invariant 2-form field is, consequently,
\be
(1,0) \oplus (0,1).
\ee
The electromagnetic field strength $F_{\mu \nu} = \partial_\mu A_\nu - \partial_\nu A_\mu$, with $n = 6$ components, transforms under this spin-1 representation. It satisfies the dynamical field equation
\be
\partial_\nu F^{\mu \nu} = 0,
\label{Afe}
\ee
which follows from the lagrangian
\be
{\mathcal L} = - \onefourth F_{\mu \nu} F^{\mu \nu}.
\ee
In terms of the potential $A_\mu$, and assuming again the Lorenz gauge (\ref{EleLoreGau}), the field equation (\ref{Afe}) reduces to the potential field equation (\ref{*}).

If an electromagnetic wave has helicity $\sigma = \pm 1$, it cannot be described by a solution of the wave equation (\ref{*}) because the electromagnetic potential $A_\mu$ lives in the representation $(1/2,1/2)$, which describes waves with helicity $\sigma = 0$. Rather, it must be identified with the solution of the wave equation (\ref{Afe}). In fact, since $F_{\mu \nu}$ lives in the representation $(0,1) \oplus (1,0)$, its solution will represent waves with helicity $\sigma = \pm 1$. Identifying the electric and magnetic components of the electromagnetic field, respectively, by
\be
E^{i} = {F}^{0 i} \quad \mbox{and} \quad
H^{i} = \onehalf \, \epsilon^{ijk} \, {F}_{j k},
\label{**88}
\ee
the field equation (\ref{Afe}) is found to be equivalent to
\be
\Box \, {\bm E} = 0 \quad \mbox{and} \quad \Box \, {\bm H} = 0,
\label{pemwaves}
\ee
where ${\bm E} = (E^i)$ and ${\bm H} = (H^i)$. Electromagnetic waves are solutions to these equations. Accordingly, they are field strength waves, not potential waves.

\subsection{A remark on photons}

As we have seen, the potential waves governed by equation (\ref{*}) have helicity $\sigma = 0$. This does not mean that photons, the quantum of the electromagnetic potential $A_\mu$, have helicity $\sigma = 0$. To understand why, let us recall that in the quantization procedure of the electromagnetic field, one starts with the definition of total angular momentum in terms of the fields $\bf E$ and $\bf H$,
\be 
\Omega^i=\int{d^3 r\left[{\bf r}\times:({\bf E}_\perp\times{\bf H}):\right]^i},
\ee
with $:\; :$ denoting normal ordering. This operator can be separated into orbital and spin parts 
\be 
\Omega^i=\Omega^i_{\rm orb} + \Omega^i_{\rm spin},
\ee
where, in terms of the vector potential, the spin part is written as
\be \label{spinop}
\Omega^i_{\rm spin}=-\int{d^3r:\frac{\partial A^j}{\partial t}\epsilon_{ijk}A^k:} \, .
\ee
Considering that the vector potential belongs to the representation $(1/2,1/2)$ of the Lorentz group, we see that the spin operator belongs to the anti-symmetric product of representations 
\[
\left[(1/2,1/2) \otimes (1/2,1/2)\right]_a ,
\]
which contains the representation $(1,0)$. This means that photons have helicity $\sigma = \pm 1$.
When the vector potential is written in terms of creation and annihilation operators $a$ and $a^\dagger$, one obtains \cite{gross}
\be \label{spinop2}
\Omega_{\rm spin}=\sum_n{\hat{z}_n\left[a^\dagger_{n,+}a_{n,+}-a^\dagger_{n,-}a_{n,-}\right]},
\ee
where $\hat{z}_n$ is an unit vector in the $+z$ direction, and $a_{n,\pm}$ is the creation operator for a photon with frequency $\omega_n$ and helicity $\sigma = \pm 1$. The fact that the solutions of the equation (\ref{*}) represent waves with helicity $\sigma = 0$ can be interpreted as a classical property of the vector potential waves, which are however not eigenstates of the spin operator (\ref{spinop2}).

\section{Massless spin-2 field}

We turn now our attention to the spin-2 representations of the Lorentz group. We follow the same strategy used in the study of the spin-1 representations.

\subsection{Potential waves}

\subsubsection{Helicity $\sigma = 0$ waves}

The first way to construct a spin-2 representation is to consider the direct product \cite{ramond}
\be
(1/2,1/2) \otimes (1/2,1/2) = [(1,1) \oplus (0,0)]_s \oplus [(0,1) \oplus (1,0)]_a,
\label{h0repre}
\ee
where the subscripts $s$ and $a$ denote, respectively, the symmetric and the anti-symmetric parts of the direct product. Such representation, with sixteen components, corresponds to a translational-valued 1-form $\varphi_\nu = \varphi^a{}_\nu P_a$, with $P_a = \partial_a$ the translation generators. Conceptually, it is equivalent to a linear perturbation of the tetrad,
\be
h^a{}_\nu = e^a{}_\nu + \varphi^a{}_\nu,
\label{TetraPertur}
\ee
where $e^a{}_\mu$ is a trivial background tetrad related to the Minkowski spacetime \cite{livro}. As such, in Cartesian coordinates it satisfies the condition $\partial_\rho e^a{}_\mu = 0$. In the so-called harmonic gauge, in which the field satisfies
\be
\partial^\nu \psi^{a}{}_\nu \equiv
\partial^\nu (\varphi^{a}{}_\nu - \onehalf \, e^{a}{}_\nu \varphi) = 0,
\label{harmo2}
\ee
where $\varphi = e_c{}^\rho \varphi^c{}_\rho$, the field equation for $\psi^a{}_\mu$ assumes the form
\be
\Box \psi^a{}_\nu = 0.
\label{h0we1}
\ee
This theory, as can be easily verified, is invariant under the gauge transformations
\be
\psi'^{a}{}_\nu = \psi^{a}{}_\nu - \partial_\nu \xi^a,
\label{gauge2}
\ee
with $\xi^a$ an arbitrary parameter. From the sixteen original components of $\psi^a{}_\mu$, the four coordinate conditions (\ref{harmo2}) and the four constraints implied by the gauge transformations (\ref{gauge2}) eliminate eight components. The invariance of the field equation (\ref{h0we1}) under local Lorentz transformations,
\be
\psi'^{a}{}_\nu = \Lambda^a{}_b \, \psi^{b}{}_\nu,
\ee
eliminates the additional 6 components of $\psi^a{}_\mu$, reducing the number of independent components to only two, as appropriate for a massless field.

In the study of gravitational waves, however, one usually assumes them to be represented by a symmetric second-rank tensor $\varphi_{\mu\nu} = \varphi_{\nu\mu}$. In this case, the representation on the right-hand side of (\ref{h0repre}) reduces to the ten components representation
\be
(1,1) \oplus (0,0),
\label{11rep}
\ee
which corresponds to the symmetric part of the direct product. Conceptually, this field is equivalent to a linear perturbation of the metric,
\be
g_{\mu \nu} = \eta_{\mu \nu} + \varphi_{\mu\nu},
\ee
with $\eta_{\mu \nu}$ the background Minkowski metric. In harmonic coordinates, in which the field satisfies the condition
\be
\partial^\nu \psi^\mu{}_\nu \equiv
\partial^\nu (\varphi^{\mu}{}_\nu - \onehalf \, \delta^{\mu}{}_\nu \varphi) = 0,
\label{harmo1}
\ee
the field equation is found to be
\be
\Box \psi_{\mu\nu} = 0.
\label{h0we}
\ee
This theory, as is well known, is invariant under the gauge transformations
\be
\psi'_{\mu \nu} = \psi_{\mu \nu} - \partial_\mu \xi_\nu - \partial_\nu \xi_\mu,
\label{gauge1}
\ee
with $\xi_\mu$ an arbitrary parameter. The four coordinate conditions (\ref{harmo1}) and the four constraints implied by the gauge invariance (\ref{gauge1}) reduce the original ten components of $\psi_{\mu\nu}$ to only two.

\subsubsection{Comparing the potential waves}

Substituting the tetrad (\ref{TetraPertur}) in the relation
\be
g_{\mu \nu} = h^a{}_\mu h^b{}_\nu \, \eta_{ab}
\ee
between the spacetime and the tangent--space metrics, we find that in Minkowski spacetime $\varphi_{\mu \nu}$ and $\varphi_{a \mu}$ are related by
\be
\varphi_{\mu\nu} = e^a{}_\mu \varphi_{a\nu} + e^a{}_\nu \varphi_{a \mu}.
\ee
Using this relation we see that their field equations, given respectively by Eqs.~(\ref{h0we}) and (\ref{h0we1}), are completely equivalent. This means that the potentials $\varphi_{\mu\nu}$ and $\varphi_{a\nu}$ describe equivalent free theories. The Feynman propagators are consequently also equivalent: whereas for $\varphi_{\mu\nu}$ it is given by
\be
D_{\mu \nu \rho \sigma}(k) = \frac{1}{2} \left(\eta_{\mu \rho} \eta_{\nu \sigma} +
\eta_{\mu \sigma} \eta_{\nu \rho} - \eta_{\mu \nu} \eta_{\rho \sigma} \right)
\left(\frac{i}{k^2 + i \varepsilon} \right),
\ee
for $\varphi_{a\nu}$ it has the form
\be
D_{a \nu b \sigma}(k) = \frac{1}{2} \left(\eta_{a b} \eta_{\nu \sigma} + 
e_{a \sigma} e_{b \nu} - e_{a \nu} e_{b \sigma} \right)
\left(\frac{i}{k^2 + i \varepsilon} \right),
\ee
where $k$ is the wave four-vector. In both cases the propagator of $\varphi_{00}$ mediates, in the static limit, an attractive gravitational potential, which is consistent with the Newtonian limit \cite{maggiore}.

When one considers the coupling the fundamental spin-2 fields $\psi_{a \nu}$ and $\psi_{\mu \nu}$ to gravitation, however, an important difference emerges. The gravitational coupling prescription for the symmetric, second--rank tensor $\psi_{\mu \nu}$, as is well known, is given by
\be
\partial_\rho \psi_{\mu \nu} \to \partial_\rho \psi_{\mu \nu} -
\Gamma^\sigma{}_{\mu \rho} \, \psi_{\sigma \nu} - \Gamma^\sigma{}_{\nu \rho} \, \psi_{\mu \sigma},
\label{GCP1}
\ee
with $\Gamma^\sigma{}_{\mu \rho}$ the Levi--Civita connection. On the other hand, owing to the fact that the algebraic index $``a$'' of the translational--valued potential $\psi_{a \nu}$ is not an ordinary vector index, but a gauge index, the coupling prescription in this case has the form
\be
\partial_\rho \psi_{a \nu} \to \partial_\rho \psi_{a \nu} -
\Gamma^\sigma{}_{\nu \rho} \, \psi_{a \sigma}.
\label{GCP2}
\ee
Although the free theories coincide, therefore, the gravitationally--coupled theories for $\psi_{\mu \nu}$ and for $\psi_{a \nu}$ will differ substantially. This difference will be important for the discussion of the last section.

\subsection{Field strength waves}

Differently from the electromagnetic field, there are two ways of constructing field strength waves for a fundamental spin-2 field. One of them has helicity $\sigma = \pm 1$ and the other has helicity $\sigma = \pm 2$. In what follows we explore in details each one of them.

\subsubsection{Helicity $\sigma = \pm 1$ waves}

Let us consider the spin-2 representation that comes from the direct product
\be
({1}/{2}, 1/2) \otimes [(1, 0) \oplus (0, 1)] =
(3/2, 1/2) \oplus (1/2, 3/2) \oplus \dots \, ,
\ee
where the dots represent additional terms that can be removed by the symmetries and constraints. By construction, it describes a translational--valued 2-form with twenty--four components. As a matter of fact, it describes the Fierz tensor ${\mathcal F}_a{}^{\mu \nu}$, which is the excitation 2-form of a fundamental spin-2 field \cite{FP}. It satisfies the Fierz equation
\be
\partial_\nu {\mathcal F}_a{}^{\mu \nu} = 0,
\label{s2Ffe}
\ee
which can be obtained by linearizing the potential form \cite{moller} of Einstein equation in the tetrad formalism.

A simpler way to approach the Fierz--Pauli formulation of a spin-2 fundamental field is to note that it is similar to the teleparallel equivalent of general relativity. In fact, the field equation (\ref{s2Ffe}) follows quite naturally from linearizing the gravitational equation of teleparallel gravity \cite{fierzTG}. In this context, the Fierz tensor is easily seen to be given by
\be
{\mathcal F}_a{}^{\mu \nu} = e_a{}^\rho \, {\mathcal K}^{\mu \nu}{}_\rho +
e_a{}^\mu \, e_b{}^\rho \, F^{b \nu}{}_\rho -
e_a{}^\nu \, e_b{}^\rho \, F^{b \mu}{}_\rho,
\label{SuperFree}
\ee
where
\be
{\mathcal K}^{\mu \nu}{}_\rho = \onehalf \left(e_a{}^\nu \, F^{a \mu}{}_\rho +
e^a{}_\rho \, F_a{}^{\mu \nu} - e_a{}^\mu \, F^{a \nu}{}_\rho \right)
\ee
is a contortion--type tensor, with
\be
F^a{}_{\mu \nu} = \partial_\mu \varphi^a{}_{\nu} - \partial_\nu \varphi^a{}_{\mu}
\label{FS}
\ee
a torsion--type tensor which represents the spin-2 field strength. The field equation (\ref{s2Ffe}) follows from the Fierz--type lagrangian
\be
{\mathcal L} = {\textstyle \frac{1}{4}} \, {\mathcal F}_a{}^{\mu \nu} \, F^a{}_{\mu \nu}.
\label{Ele}
\ee

Analogously to the electromagnetic case, the gravitoelectric and the gravitomagnetic components of the gravitational field can be defined, respectively, by \cite{vanessa}
\be
E^{a i} = {\mathcal F}^{a 0 i} \quad \mbox{and} \quad
H^{a i} = \onehalf \, \epsilon^{ijk} \, {\mathcal F}^{a}{}_{j k}.
\label{**}
\ee
Equation (\ref{s2Ffe}) is then found to be equivalent to
\be
\Box \, {\bm E^a} = 0 \quad \mbox{and} \quad \Box \, {\bm H^a} = 0,
\label{h1wave}
\ee
where ${\bm E^a} = (E^{a i})$ and ${\bm H^a} = (H^{a i})$. Owing to the fact that ${\bm E^a}$ and ${\bm H^a}$ are components of ${\mathcal F}^{a\mu \nu}$, which is a field that lives in the representation $(1/2, 3/2) \oplus (3/2, 1/2)$ of the Lorentz group, the solutions to the equations (\ref{h1wave}) represent spin-2 waves with helicity $\sigma = \pm 1$.

\subsubsection{Helicity $\sigma = \pm 2$ waves}

Another spin-2 representation can be constructed by considering the direct product
\be
[(0,1) \oplus (1,0)] \otimes [(0,1) \oplus (1,0)] = (0,2) \oplus (2,0) \oplus \dots \, ,
\ee
where again the dots represent additional terms that can be removed by the symmetries and constraints. As is well known, this representation describes a fourth--rank tensor which is antisymmetric within each pair of indices, and symmetric between the pairs. It represents actually the twenty components of a Riemann--like tensor ${\mathcal R}^\rho{}_{\lambda \mu \nu}$ constructed out of the second derivatives of the symmetric field $\varphi_{\mu\nu}$,
\be
{\mathcal R}^\rho{}_{\lambda \mu \nu} = \partial_\mu \gamma^\rho{}_{\lambda \nu} -
\partial_\nu \gamma^\rho{}_{\lambda \mu},
\ee
with
\be
\gamma^\rho{}_{\lambda \nu} = \onehalf \, \eta^{\rho \sigma} \left(
\partial_\lambda \varphi_{\nu \sigma} + \partial_\nu \varphi_{\lambda \sigma} -
\partial_\sigma \varphi_{\lambda \nu} \right).
\ee
Defining the Ricci--type tensor ${\mathcal R}_{\lambda\nu} = {\mathcal R}^\rho{}_{\lambda \rho \nu}$, the curvature wave is found to satisfy the Einstein type equation
\be
{\mathcal R}_{\lambda\nu} = 0,
\label{CurSol}
\ee
which in harmonic coordinates reduces to the potential field equation (\ref{h0we}), that is,
\be
\Box \psi_{\mu\nu} = 0.
\label{h0weBis}
\ee
As is well known, it follows from the Einstein--Hilbert type lagrangian
\be
{\mathcal L} = - {\mathcal R} \equiv - \eta^{\lambda\nu} \, {\mathcal R}_{\lambda\nu}.
\label{EHL}
\ee
Due to the fact that ${\mathcal R}^\rho{}_{\lambda \mu \nu}$ lives in the representation $(0,2) \oplus (2,0)$ of the Lorentz group, the solution of the equation (\ref{CurSol}) describes spin-2 waves with helicity $\sigma = \pm 2$.

Up to a surface term, the lagrangian (\ref{EHL}) is equivalent to the Fierz--type lagrangian (\ref{Ele}). The presence of this surface term is related to the fact that, whereas the lagrangian (\ref{Ele}) depends on $\varphi_{a \nu}$ and on its first derivative, lagrangian (\ref{EHL}) includes also second derivatives of $\varphi_{\mu \nu}$, which reduces however to a divergence. For the same reason, the field equation (\ref{CurSol}) does not involve derivatives of the curvature tensor. The curvature--like tensor ${\mathcal R}^\rho{}_{\lambda \mu \nu}$, therefore, is a derived geometrical entity which does not have its own dynamics: its propagation is simply a consequence of the dynamics of the potential $\varphi_{\mu \nu}$. This is different from the electromagnetic case, whose field strength $F^{\mu \nu}$ satisfies a dynamical equation. It is also different from the Fierz--Pauli formulation of gravity, whose excitation 2-form ${\mathcal F}_a{}^{\mu \nu}$ satisfies a dynamical equation.

\section{Conclusions}

Gravitational waves are believed to be waves with helicity $\sigma = \pm 2$. This idea is related to the fact that they are described by the curvature--like tensor ${\mathcal R}^{\rho}{}_{\lambda\mu\nu}$, which is a field that lives in the representation $(0,2) \oplus (2,0)$ of the Lorentz group. However, this is not the only possibility. In fact, in the Fierz--Pauli approach to gravitation (or equivalently, in the teleparallel approach), gravitational waves are described by the field excitation ${\mathcal F}^{a\mu \nu}$, which is a field that lives in the representation $(1/2, 3/2) \oplus (3/2, 1/2)$ of the Lorentz group. As such, it describes waves with helicity $\sigma = \pm 1$. Whatever the case, it is important to remark that, similarly to the electromagnetic waves, gravitational waves are not {\em potential} waves, but {\em field excitation} waves. The question then arises: is there any reason to neglect waves with helicity $\sigma = \pm 1$ in favor of waves with helicity $\sigma = \pm 2$? From the point of view of representation theory, there is none: both representations are well defined Lorentz representations, and consequently able to describe physical waves.

If some reasons exist, they might then be related to the different aspects of the physical theories describing them. For example, all attempts to couple the symmetric potential $\psi_{\mu \nu}$ to gravitation have led to inconsistent theories \cite{AraDeser}. More specifically, when the gravitational coupling prescription (\ref{GCP1}) is applied to the free field equation (\ref{h0weBis}), the resulting theory is no longer gauge invariant, and consequently not all spurious components of the potential can be eliminated. Furthermore, owing to the fact that the commutator of covariant derivatives is proportional to the Ricci curvature tensor, unphysical constraints on the spacetime geometry show up. Addition of non-minimal coupling terms has also been shown not to cure the problem \cite{DesHen}.
On the other hand, when a spin-2 potential is assumed to be represented by a 1-form assuming values in the Lie algebra of the translation group, and the coupling prescription (\ref{GCP2}) is applied to the field equation (\ref{s2Ffe}), a sound and consistent gravitationally-coupled spin-2 field theory emerges, which is quite similar to the gravitationally-coupled electromagnetic theory. It preserves the gauge invariance of the free theory, and no unphysical constraints are imposed on the geometry of spacetime \cite{cqg1}.

Put together, the above results can be considered a compelling evidence favoring the Fierz--Pauli approach (or equivalently, the teleparallel approach) to a fundamental spin-2 field, and consequently the interpretation of spin-2 waves as field excitation ${\mathcal F}^{a\mu \nu}$ waves with helicity $\sigma = \pm 1$ waves.\footnote{In a different context, gravitational waves with helicity $\sigma = \pm 1$ have been discussed in Ref.~\cite{spin1GW}.} However, since the concept of helicity can only be defined at the linear level \cite{wald}, and taking into account that gravitational waves are essentially non--linear \cite{GW}, it is not clear whether the concepts related to a fundamental spin-2 field can be immediately extended to gravitational waves. Even if the linear approximation to the gravitational waves theory makes physical sense, it can always be corrected by higher--order terms, in which case the very notion of helicity would be lost.

\section*{Acknowledgments}
The authors would like to thank Y. Obukhov for useful discussions, and an anonymous referee for useful comments and suggestions. They would like to thank also FAPESP, CNPq and CAPES for partial financial support.

\end{document}